\documentclass[prl,twocolumn,secnumarabic, showkeys]{revtex4}
\usepackage{bm}
\usepackage{amssymb}
\usepackage{mathtools}
\usepackage{amsmath}
\usepackage{relsize}
\usepackage{hyperref}
\usepackage[mathscr]{euscript}
\usepackage{enumitem}
\usepackage{natbib}
\usepackage{etoolbox}

\makeatletter
\pretocmd\frontmatter@keys@format{\addvspace{11\p@}}{}{}
\makeatother

\begin{document}
\title{The status of geometry and matter in re-interpreted WdW equation}
\author{Avadhut V. Purohit}
\email{avdhoot.purohit@gmail.com\\}
\affiliation{Visvesvaraya National Institute of Technology, S Ambazari Rd, Ambazari, Nagpur, India 440010}
\date{\today}

\begin{abstract}
  I have shown that the field defined by the Wheeler-DeWitt equation for \textit{pure gravity} is neither a standard gravitational field nor the field representing a particular universe. The theory offers a unified description of geometry and matter, with geometry being fundamental. The quantum theory possesses gravitational decoherence when the signature of $R^{(3)}$ changes. The quantum theory resolves singularities dynamically. Application to the FLRW $\kappa=0$ shows the creation of local geometries during quantum evolution. The 3-metric gets modified near the classical singularity in the case of the Schwarzschild geometry.
\end{abstract}


\maketitle
\section{Introduction}
 In search of the most fundamental theory, mostly two schools of thought are prevalent. One school of thought thinks that the quantum theory is fundamental. Because small objects together make a big object. Therefore, gravity gets quantized at a small scale. Often, the existence of singularities in the classical theory indicates limitations of the theory. The theories such as the loop quantum gravity and the Wheeler-DeWitt theory take this approach. The quantization of a relativistic particle shows that the quantization by raising $E\rightarrow i\hbar \partial_t$ and $\vec{p}\rightarrow -i\hbar \vec{\nabla}$ to operators on the Hilbert space faces several mathematical and conceptual issues. By taking a leaf out of the quantum field theories; \citep{ThirdQuantization} and others tried to re-interpret the Wheeler-DeWitt equation. But this approach also has severe issues. Although the gravitational and matter fields are fields defined over a 4-dimensional manifold, the matter fields involve re-interpretation. But the gravitational field is a generalization of the special theory of relativity. It does not involve re-interpretation. When you re-interpret the full Wheeler-DeWitt equation, the gravitational field and matter fields are not on equal footing. Therefore, such quantization is dubious (refer \citep{kucher} and \citep{KieferBook} for more details).\\
 \hspace*{3mm}Another school thinks that gravity being a dynamical theory of spacetime itself, is more fundamental. The quantum theory would receive modifications in the case of a gravitational scenario. Recently there have been new developments \citep{penrose} in this approach. Everything else ceases to exist in the absence of background spacetime. Apart from the search for a unified particle theory that included gravity as a natural ingredient, theoreticians pursued a separate line of investigation schooled primarily in general relativity and topology. Clifford presented his paper ``On the Space Theory of Matter'' \citep{clifford} and Einstein gave substance to this line of inquiry. He writes, ``The material particle has no place as a fundamental concept in field theory. Even electrodynamics is not complete for this reason. Gravity as a field theory must also deny a preferred status to matter.'' John Wheeler also attempted to build such a theory. He writes, ``What else is there out of which to build a particle except for geometry (spacetime) itself?'' Canonical theories such as the Wheeler-DeWitt theory quantize gravity. But they do not describe geometric matter (\citep{odenwald}, page 8). Unfortunately, even after having a philosophical background, there has been no success in this direction so far.\\
\hspace*{3mm}In this paper, I analyse the Wheeler-DeWitt equation for \textit{pure gravity} in the light of standard quantum field theories. Because the field is defined only over the space of 3-metric can be re-interpreted without the issue discussed above. The field defined satisfies ADM constraints for pure gravity. Therefore, one would interpret that the field $\Phi$ is a pure gravitational field. But I observe that even gauge fields obey ADM constraints for pure gravity. I also observe that these fields have non-trivial stress tensors. Whereas the stress tensor for pure gravitational field is $R_{\mu\nu}-\frac{1}{2}g_{\mu\nu}R = 0$. The quadratic coupling always remains non-negative regardless of the signature of $R^{(3)}$. I also observe that the higher-order couplings with $\Phi\sim e^{i q_{ab}P^{ab}}$ allow us to interpret it as a matter-like term. The other interpretation is that the field $\Phi$ describes a particular Universe. I observe that such interpretation faces problems due to the interaction between different fields. It shows that neither of the interpretation is true. The field $\Phi$ is a unified description of the gravity and scalar matter. \\
\hspace*{3mm}The re-interpretation partly modifies both the theories, the quantum theory as well as the classical gravity. On the quantization of the field, we get the geometric quantum corresponding to the field. There is no quantum of gravity or graviton. With the curvature's signature-dependent decoherence, the vector-valued creation and annihilation operators show gravitational effects on the nature of the quantum theory. Whereas the dynamical singularity resolution shows a modification of the classical gravity. The quantum with a particular 3-metric has a definite energy. But it does not have well-defined momentum. The theory applied to the FLRW $\kappa=0$ Universe creates non-trivial local geometric quanta. The theory applied to the Schwarzschild geometry shows the existence of the Planck scale black hole. The sense of distance arises from the interaction between fields.\\
\hspace*{3mm}The re-interpretation connects the requirements of the correct quantum theory and expectations of a general relativistic school of thought.
\section{The scalar field}
The Wheeler-DeWitt equation written in the DeWitt's coordinates 
\begin{eqnarray}\label{FieldEqn}
  \left(\frac{1}{\sqrt{-G}}\partial_\mu \sqrt{-G}G^{\mu\nu}\partial_\nu + \mu^2\right) \Phi = 0
\end{eqnarray}
 is interpreted as a classical field equation. The field $\Phi$ is defined over the superspace $\zeta^{\mu} \coloneqq \left( \zeta, \zeta^{A} \right)$ with $A=1,2,3,4,5$. The DeWitt super-metric borrowed from \citep{DeWitt}.
\begin{eqnarray} 
& G_{\mu\nu} \coloneqq
  \begin{pmatrix}
    1  &0 \\
    0  &-\frac{3 \zeta^{2}}{32}\bar{G}_{AB}
  \end{pmatrix}
  &\bar{G}_{AB} \coloneqq \text{Tr}\left(\mathbf{q}^{-1}\frac{\partial \mathbf{q}} {\partial\zeta^A}\mathbf{q}^{-1}\frac{\partial \mathbf{q}}{\partial\zeta^B}\right) \\ \nonumber
    & \zeta \coloneqq \left( \frac{32}{3}\right)^\frac{1}{2} \left( \text{det }q_{ab} \right)^\frac{1}{4} 
    &\sqrt{-G}\coloneqq \sqrt{-\text{det }G_{\mu\nu}}
\end{eqnarray}
Here, $\mathbf{q} \coloneqq q_{ab}$. $\bar{G}_{AB}$ is symmetric super-metric on 5D manifold $M$ identified with $SL(3, \mathbb{R})/SO(3, \mathbb{R})$ (refer \citep{DeWitt} and \citep{Giulini} for more discussion on the geometry of superspace). As discussed by DeWitt \citep{DeWitt} in appendix A, $\zeta^A$ are new orthogonal coordinates chosen from components of the 3-metric as they act as ``good'' coordinates. The coupling function is defined as
\begin{equation}\label{coupling}
    \mu^2 \left(\zeta,\zeta^{A}\right)\coloneqq  \frac{3\zeta^{2}}{32} R^{(3)}
\end{equation}
The field $\Phi$ is functional over the space of 3-metric only. Hence, $\frac{\partial \Phi}{\partial q_{ab}}$ is also a functional over 3-metric. Trivially, it satisfies the diffeomorphism constraints
\begin{equation}
  D_{a}P^{ab} \Rightarrow D_{a}\frac{\partial \Phi}{\partial q_{ab}} \approx 0
\end{equation}
The action functional for the geometric scalar field that gives field equation (\ref{FieldEqn}) is assumed to have the following form.
\begin{eqnarray}\label{ScalarAction}
  \mathcal{A}_{\Phi} \coloneqq \int \, \mathscr{D}\zeta \frac{\sqrt{-G}}{2} \left( \partial_{\mu}\Phi G^{\mu\nu}\partial_{\nu}\Phi - \mu^2 \Phi^{2} \right)
\end{eqnarray}
$\mathscr{D}\zeta$ is suitable measure over 6D manifold. The ADM theory does not have an issue with operator ordering. But for the classical geometric scalar field, different combinations of $P^{ab}$ and $q_{ab}$ gives non-equivalent field equations. I have taken the combination of field variables with a consistent self-adjoint extension. In other words, the combination allows the Hamiltonian operator to be self-adjoint. On single spacetime-like interpretation $\Phi \sim e^{i q_{\mu\nu}P^{\mu\nu}}$, ADM Hamiltonian constraints for pure gravity in the DeWitt coordinates (5.20, \cite{DeWitt})
\begin{equation}\label{HamiltonianConstraints}
  P_0^2 - \frac{32}{3 \zeta^2}\bar{G}^{AB}P_A P_B + \frac{3\zeta^{2}}{32} R^{(3)}\approx 0
\end{equation}
get recovered.\\
\hspace*{3mm}Invariance of an action under variation $\zeta^{\mu} \rightarrow \zeta^\mu + \delta \zeta^\mu$ gives stress tensor.
\begin{eqnarray} \label{StressTensor}
 T^{\nu}_{\mu} \coloneqq \frac{\partial \mathcal{L}}{\partial \left( \frac{\partial \Phi}{\partial \zeta^{\nu}}\right)}\frac{\partial \Phi}{\partial \zeta^{\mu}} - \mathcal{L}\delta_{\mu}^{\nu}
\end{eqnarray}
For now, assume $\zeta$ as time and perform Legendre transformation to get the Hamiltonian
\begin{eqnarray}\label{Hamiltonian}
    & \scriptstyle \Pi_\Phi \coloneqq \frac{\partial \left(\sqrt{-G}\mathcal{L}\right)}{\partial \frac{\partial \Phi}{\partial \zeta}} = \sqrt{-G}\frac{\partial \Phi}{\partial \zeta} \\
  &\scriptstyle \mathbf{H}_{\Phi} = \int \, \mathscr{D}\zeta^{A} \, \frac{1}{2}\left( \frac{\Pi_\Phi^{2}}{\sqrt{-G}} + \frac{32}{3\zeta^{2}}\sqrt{-G}\frac{\partial \Phi}{\partial \zeta^{A}}\bar{G}^{AB}\frac{\partial \Phi}{\partial \zeta^{B}} + \sqrt{-G}\mu^2 \Phi^{2} \right)
\end{eqnarray}
For $\mu^2 < 0$, the field is self-coupled, i.e. it has $h \Phi^{4}$ with some $h>0$. The Hamiltonian shows that the quadratic coupling will be non-negative regardless of the signature of the 3-Ricci curvature scalar and justifies the use of $\zeta$ as the time for the geometric scalar field. 
 On a single-geometric interpretation, we get
\begin{equation}
  P_0^2 - \frac{32}{3 \zeta^2}\bar{G}^{AB}P_A P_B + \frac{3\zeta^{2}}{32} R^{(3)}\approx h
\end{equation}
The coupling parameter $h$ is free and does not necessarily depend on the geometry of spacetime. From ADM theoretic viewpoint, $h\neq 0$ represents the matter field. For $\mu^2 < 0$, the minima lies at $\Phi_0 = \pm \sqrt{\frac{-2\mu^2}{\lambda}}$. The field rolls down to get positive quadratic coupling. $h$ appears due to field theoretic reason and gravity guides us to interpret the coupling term $h$ as a matter-like term. \\
\hspace*{3mm}If we look at the Green's function for the Wheeler-DeWitt operator \ref{FieldEqn}, in the limit $\zeta\rightarrow\infty$, the middle spatial terms become negligible and we effectively get
\begin{equation}
	\left(\frac{\partial^2}{\partial\zeta^2}+\frac{1}{\zeta}\frac{\partial}{\partial\zeta}+\mu^2\right) Q\left(\zeta, \zeta^\prime \right) = \delta\left(\zeta-\zeta^\prime \right)
\end{equation}
The Green's function for non-zero constant $\mu^2$ is
\begin{equation}
	Q = \frac{\pi}{2}\theta\left(\zeta - \zeta^\prime \right) \zeta^\prime \left( Y_0 \left(\mu\zeta\right)J_0 \left(\mu\zeta^\prime\right) -J_0 \left(\mu\zeta\right)Y_0 \left(\mu\zeta^\prime\right) \right)
\end{equation}
and for $\mu^2=0$, it is
\begin{equation}
	Q = \theta\left(\zeta - \zeta^\prime \right)\ln\left(\frac{\zeta}{\zeta^\prime}\right)
\end{equation}
The Heaviside function $\theta\left(\zeta - \zeta^\prime \right)$ is zero for $\zeta < \zeta^\prime$ and 1 for $\zeta > \zeta^\prime$. Hence, the signal propagates forward in $\zeta$. It shows that $\zeta$ acts as time for the field $\Phi\left(\zeta^\mu\right)$. \\ 
\hspace*{5mm}Note: \textit{The Ricci curvature scalar $R^{(3)}$ and cosmological constant $\Lambda$ are not on equal footing. $R^{(3)}$ appears in quadratic coupling, whereas $\Lambda$ contributes to the vacuum}.
\subsection{Features}\label{features}
\hspace*{3mm} The field $\Phi (q_{ab})$ is identified with an intrinsic property $\mu$ and defined over $q_{ab}$ which is the solution to intrinsic curvature $R^{(3)}$. The existence of dynamical background shows that the field is gravitational.\\
\hspace*{3mm} Even when the 3-geometry have positive curvature, quadratic coupling remains positive. This shows resemblance of the field with the standard matter fields.\\
\hspace*{3mm} The field $\Phi$ can also have charge. 
  \begin{equation}\label{complex}
    \mathcal{A}_{\text{complex}} \coloneqq \int \, \mathscr{D}\zeta \frac{\sqrt{-G}}{2} \left( \partial_{\mu}\Phi^\star G^{\mu\nu}\partial_{\nu}\Phi - \mu^2 \Phi^\star \Phi \right)
  \end{equation}
  The pure gravity on other hand, does not have opposite charges. This is the property that resembles charged matter.\\
\hspace*{3mm} The existence of $\mu^2 = 0$ or $R^{(3)} = 0$ does not necessarily mean line element is zero.
  \begin{align}\label{lineElement}
    &ds^2 = d\zeta^2 - \frac{3\zeta^2}{32}\bar{G}_{ij} dx^i dx^j
    &\bar{G}_{ij} \coloneqq \bar{G}_{AB}\frac{\partial \zeta^A}{\partial x^i}\frac{\partial \zeta^B}{\partial x^j}
  \end{align}
  For the 3-metric $q_{ab} \coloneqq q(t)\, \text{diag}\left( f(r), r^2, r^2 \sin^2 \theta \right)$, the line element is given as follows.
  \begin{equation}
    ds^2 = G^{abcd}dq_{ab}dq_{cd} = 
    \begin{pmatrix}
      dq &df  
    \end{pmatrix}\begin{pmatrix}
      \frac{-6}{f^\frac{1}{2} q^\frac{1}{2}}  &-2\frac{q^\frac{1}{2}}{f^\frac{3}{2}}\\
      -2\frac{q^\frac{1}{2}}{f^\frac{3}{2}}  &0
    \end{pmatrix}\begin{pmatrix}
      dq \\
      df  
    \end{pmatrix}
  \end{equation}
  The \textit{distance} in the superspace, in the large $q$ limit, $ds^2 = -4\frac{q^\frac{1}{2} dq\,df}{f^\frac{3}{2}}$ increases with $q$. It shows cosmological expansion. The distance decreases with increase in $f$, showing attraction between two objects. This distance is actually the distance between two 3-metrics (refer \citep{DeWitt}).\\
\hspace*{3mm} The energy of the field $\Phi$ is always well-defined. Gravitational field may not always have time-like Killing vector field. Therefore, defining energy in the general relativity is not straight forward.
\section{Re-interpretation and gauge invariance}
A famous experiment performed in 1975 by Colella, Overhauser and Werner did confirm that quantum mechanics respect the principle of equivalence (page 11, \citep{penrose}). In general relativization of the quantum theory, wave functions get non-trivial phase differences. In the ADM theory, the set of lapse functions $N$ and shift vector $N^a$ identify the frame of reference.\\
\hspace*{3mm}The field $\Phi$ contains information about reference frame which can be easily seen by single-geometric interpretation $\Phi \sim e^{\pm i P^{ab}q_{ab}}$ leading to (\ref{HamiltonianConstraints}). Here, $P^{ab}$ contains information about the lapse function and shift-vector. The complex scalar field $\Phi$ is not invariant under transformation $\Phi \rightarrow e^{i \alpha \left(\zeta^\mu\right)}\Phi$. It requires the field $A_\mu$ that makes (\ref{complex}) a gauge invariant. This theory fully respects the equivalence principle.
\begin{eqnarray}
  &\mathcal{L}_{\text{complex}} = \frac{1}{2}\left( D_\mu\Phi\right)^\star G^{\mu\nu}\left(D_\nu\Phi\right) - \frac{\mu^2}{2} \Phi^\star \Phi \\
  & D_\mu \coloneqq \partial_{\mu} - i \alpha A_\mu
\end{eqnarray}
The third quantized theories interpret $\Phi$ as a description of a particular universe. But we can see that the Lagrangian describes interaction between charged field with its gauge field. Here, different universes interacting with each other destroys very definition of the universe. Since the third quantized theories include matter fields in the superspace, the field $\Phi$ cannot be interpreted as a matter field as well. For the interpretation presented in this paper, there is no such problem. Because the superspace is defined only over the space of the 3-metric. \\
\hspace*{3mm}The action functional for field $A_\mu$ assumed to have the following form.
\begin{eqnarray}
    & \mathcal{A}_{\text{vector}} \coloneqq - \frac{1}{4} \int \mathscr{D}\zeta \sqrt{-G} F_{\mu\nu}F^{\mu\nu} \\ \nonumber 
    & F^{\mu\nu} \coloneqq G^{\mu\rho}G^{\nu\sigma}F_{\rho\sigma} \hspace*{2mm}\text{and}\hspace*{2mm}  F_{\mu\nu} \coloneqq \partial_{\mu}A_{\nu} - \partial_{\nu}A_{\mu}
\end{eqnarray}
$F_{\mu\nu}$ is completely anti-symmetric tensor and therefore satisfies Bianchi identity. Field equations in presence of source $J^\mu$ and in the gauge selected above are obtained as
\begin{equation}\label{GaugeFieldEquation}
    \partial_\mu F^{\mu\nu} + \frac{F^{\mu\nu}}{\sqrt{-G}}\partial_\mu \left(\sqrt{-G}\right) = J^{\nu}
\end{equation}
In addition to other sources $J^\nu$, the second geometric term also acts as source for the gauge field. In absence of source, fields $B_{A} \coloneqq \epsilon_{ABC}F_{BC}$ as well as $E_{A} \coloneqq F_{0A}$ satisfy Wheeler-DeWitt equation.
\begin{equation}
    \frac{1}{\sqrt{-G}} \partial_\mu \left(\sqrt{-G} G^{\mu\nu} \partial_\nu\right) \begin{pmatrix}
        E_C\\ B_C
    \end{pmatrix} = 0
\end{equation}
In asymptotic flat limit (i.e. $\sqrt{-G}\approx 1$), the field equations become $\partial_\mu F^{\mu\nu} = J^\nu$.\\
\hspace*{3mm}The field $A_\mu$ also satisfies the ADM constraints for pure gravity and has a non-trivial stress tensor. If $\Phi$ is interpreted as a pure gravitational field, then the field $A_\mu$ would also have to be interpreted as a pure gravitational field. But vector-valued gravitational field $A_\mu$ is a disaster. It shows that fields $\Phi$ and $A_\mu$ are not pure gravitational fields, at least not the standard gravitational fields.
\section{Quantization}
\hspace*{3mm}I define vector-valued annihilation and creation operators
\begin{eqnarray}
  & a_{A} \coloneqq \frac{(-G)^{\frac{1}{4}}}{\sqrt{2}} \biggl( \frac{\Pi}{(-G)^{\frac{1}{2}}} n_{A} + i \sqrt{\frac{32}{3\zeta^{2}}}\frac{\partial \Phi}{\partial \zeta^{A}} + i \omega \Phi n_{A} \biggr)  \\ \nonumber
  & a^{\dagger}_{B} \coloneqq  \frac{(-G)^{\frac{1}{4}}}{\sqrt{2}} \biggl( \frac{\Pi}{(-G)^{\frac{1}{2}}} n_{B} - i\sqrt{\frac{32}{3\zeta^{2}}}\frac{\partial \Phi}{\partial \zeta^{B}} - i \omega \Phi n_{B} \biggr) 
\end{eqnarray}
$n^{A}\coloneqq\frac{\zeta^A}{\sqrt{\bar{G}_{AB}\zeta^A \zeta^B}}$, $(-G)^{\frac{1}{4}} \coloneqq \left( -\text{det } G_{\mu\nu}\right)^\frac{1}{4}$ and $\omega \in \mathbb{R}$ is chosen as the solution to following Riccati equation
\begin{eqnarray} \label{Riccati}
 \sqrt{-G} \omega^{2} - \frac{\partial }{\partial \zeta^{C}} \left( \omega \sqrt{-G} \sqrt{\frac{32}{3\zeta^{2}}} \, n_{A}\bar{G}^{AC} \right) = \sqrt{-G}  \mu^2 
\end{eqnarray}
The equation should be solved using `correct' boundary conditions. Such a solution is unique. Examples of such boundary conditions are
\begin{itemize}
  \item FLRW $\kappa=0$ model: $\omega$ that makes the spectrum of the Hamiltonian operator continuous in the limit $q(t) \rightarrow \infty$.
  \item Schwarzschild spacetime: $\omega$ that makes the spectrum of the Hamiltonian operator continuous in the limit $q_{ab}\rightarrow\eta_{ab}$ with $\eta_{ab}$ being flat 3-metric. 
\end{itemize}
Computing the non-trivial commutator using property of the Dirac delta function $f(x)\delta^\prime (x) = - f^\prime (x)\delta (x)$ for $f(x)\neq \text{constant}$, we get
\begin{eqnarray}\label{commutator}
  \scriptstyle \left[ a, a^{\dagger} \right] = \frac{\varepsilon_{Pl}}{2} \left( \sqrt{\frac{32}{3\zeta^{2}}} \partial_C  n^C - \omega \right) \delta \left( \vec{\zeta}, {\vec{\zeta^{\prime}}} \right) = \beta \left( \zeta, \zeta^A\right) \delta \left( \vec{\zeta}, {\vec{\zeta^{\prime}}} \right) 
\end{eqnarray}
There is a conserved quantity corresponds to $\zeta$ that I call it as energy. I introduce $\varepsilon_{Pl}$ instead of $\hbar$ to maintain $\zeta$ dimensionless. Because here, I interpret the theory relative to the ADM theory. If we interpret $\zeta$ as time in seconds without refering to the ADM theory, we can replace $\varepsilon_{Pl}\rightarrow \hbar$. The above commutator was possible because the inverse metric $\bar{G}^{AB}$ is symmetric. These identities allow us to write the Hamiltonian operator in the discrete space. (i.e. $\int \mathscr{D}\zeta^A\rightarrow\sum_{\zeta^A}$)
\begin{eqnarray}
  & \mathbf{H}_{\Phi} = \sum_{\zeta^{A}} a^{\dagger}_{A}\bar{G}^{AB}a_{B} \\ \nonumber
  & + \frac{\delta(0)}{2}\varepsilon_{Pl}\sum_{\zeta^{A}} \left( \sqrt{\frac{32}{3\zeta^{2}}} \partial_C n^C - \omega \right) 
\end{eqnarray}
The second term is the vacuum term. The quantum vacuum is a sea of constantly creating and annihilating geometries. Discarding this term and writing the Hamiltonian operator in terms of number operator $\hat{\textbf{n}}= \sum_A \hat{\textbf{n}}_A$ with ${a^\dagger}^A a_A \coloneqq \left(\sqrt{\frac{32}{3\zeta^{2}}} \partial_C n^C - \omega \right) \hat{\textbf{n}}_A$
\begin{eqnarray} \label{HamiltonianOperator}
  \hat{\mathbf{H}}_{\Phi} = \varepsilon_{Pl} \sum_{\zeta^{A}} \left| \sqrt{\frac{32}{3\zeta^{2}}} \partial_C n^C -  \omega \right| \hat{\textbf{n}}
\end{eqnarray}
The appearance of the differential equation (\ref{Riccati}) is not surprising. It is a consequence of using coordinate-space for quantization. The scalar quantum of the Klein-Gordon field satisfies $\omega^2 = k^2 + m^2$. Similarly, the quantum of the geometric scalar field follows $frequency = \scriptstyle \sqrt{\frac{32}{3\zeta^{2}}} \partial_C n^C - \omega $ with $\omega$ being solution to (\ref{Riccati}). $\Phi$ has a single degree of freedom. Therefore the quantum is scalar. $\Pi$ is a collection of creation and destruction operators. But $\Phi$ depends non-linearly on creation and annihilation operators.\\
\hspace*{3mm}The momentum operator defined using stress tensor
\begin{equation}
  \hat{\mathbf{P}}^{C} = - \frac{32}{3\zeta^2} \sum_{\zeta^{A}}\, \left( \bar{G}^{CB} \frac{\partial \Phi}{\partial \zeta^{B}}\right)\Pi
\end{equation}
does not share eigenstates with the Hamiltonian operator. This is because $\Phi$ depends non-linearly on creation and annihilation operators. Therefore the quantum with a particular 3-metric does not have well-defined momentum at the quantum level.
\subsection{Application}
If the spacetime is spherically symmetric with the 3-metric $q_{ab} \coloneqq q(t)\text{ diag}\left(f(r), 1 ,1\right)$ with ADM coordinates $dr, rd\theta$ and $r\sin\theta \, d\phi$ chosen keeping dimensionality in mind. Then, DeWitt supermetric becomes $G_{\mu\nu}\coloneqq \textbf{diag }\left(1, -\frac{3 \zeta^{2}}{32 f^2}\right)$ with $\zeta^A \coloneqq f(r) = f$ being one dimensional. The determinant of supermetric $-\text{det }G_{\mu\nu}= \frac{3 \zeta^{2}}{32 f^2} $ , $\bar{G}_{AB}=\frac{1}{f^2}$ and $n^A = f$. That implies $\partial_C n^C =1$. Then, the Hamiltonian becomes
\begin{equation}
  \mathbf{H}_\Phi = \varepsilon_{Pl} \sum_f \left| \sqrt{\frac{32}{3}}\frac{1}{\zeta} -  \omega \right| \hat{\mathbf{n}} 
\end{equation}
The $\omega$ is a solution to the following Riccati equation.
\begin{equation}
  \omega^2 - \sqrt{\frac{32}{3}} \frac{f}{\zeta} \frac{\partial \omega}{\partial f} = \mu^2
\end{equation}
\subsubsection{Spatially flat FLRW universe}
For the spatially flat spacetime $\mu^2 = 0$, I choose trivial solution $\omega = 0$. The Hamiltonian in such case becomes
\begin{equation} \label{UniverseH}
  \hat{\mathbf{H}}_\Phi =  \sqrt{\frac{32}{3}}\frac{\varepsilon_{Pl}}{\zeta} \, \hat{\mathbf{n}}= \frac{\varepsilon_{Pl}}{a^\frac{3}{2}} \, \hat{\mathbf{n}}
\end{equation}
Here, $a$ represents the scale factor. The Hamiltonian spectrum becomes continuous in the limit $\zeta\rightarrow\infty$ that justifies selected trivial solution $\omega = 0$. The frequency (or energy) of the quantum gets redshifted (proportional to $\frac{1}{\zeta}$) with time. There exist a conserved quantity corresponding to $\zeta$. I call it an \textit{energy}. If we assume the Universe as a collection of $n$ identical quanta, then the total energy of the Universe is $E_U \approx \varepsilon_{Pl}  \frac{n}{\zeta_{min}}$. The finiteness of $E_U$ results in the existence of finite non-zero $\zeta=\zeta_{min}$.  The energy of a quantum in a particular state decreases with time $\zeta$. We need to create more quanta to conserve the total energy. But at a given time every quantum has the same energy. That means we cannot conserve the total energy in this way. One of the possible way to conserve the energy is by creating local geometries. That is, by introducing $f\neq 1$ with $R^{(3)}=0$. An individual quantum radiates energy.
\begin{equation}
  \frac{\varepsilon_{Pl}}{\zeta_{min}} \xrightarrow{\zeta \text{ evolution}}  \frac{\varepsilon_{Pl}}{\zeta_{min} +\Delta\zeta} + \varepsilon \left(\zeta , f \right)
\end{equation}
Here, $\varepsilon \left(\zeta , f\right)$ indicate created quanta in the evolution. This is locally allowed by geometries such as the Schwarzschild geometry where, $R^{(3)}=0$ but $f\neq 1$. It shows that, even though the metric in the beginning is $\zeta$ dependent only, the geometric quantum theory naturally introduces local variations of the 3-metric to conserve the total energy of the Universe.
\subsubsection{Schwarzschild geometry}
The Schwarzschild geometry written in the isotropic radial coordinates as
\begin{align}
	&q_{ab} \coloneqq f\, \text{diag}\left( 1, r^2, r^2 \sin^2\theta \right), &f=\left( 1 + \frac{M}{2r}\right)^4
\end{align}
Here, I set $\zeta =1$ for the static geometry. The Hamiltonian spectrum in this situation is obtained by taking trivial solution to the Riccati equation as it gives the correct quantum theory.
\begin{equation}
  \hat{\mathbf{H}}_\Phi \approx 3.266\, \varepsilon_{Pl}\, \hat{\mathbf{n}} \sum_f 1 
\end{equation}
As $f(r) \in \left(1, \infty \right)$, integrating $f$ from 1 to $f_{max}$, we get 
\begin{equation}
  \hat{\mathbf{H}}_\Phi \approx 3.266 \,  \varepsilon_{Pl}\left( f_{max}- 1 \right)\hat{\mathbf{n}}
\end{equation}
The energy of a quantum state increases with $f$. The energy spectrum shows area quantization, as $\sqrt{f}$ is a dimensionless length. The black hole with energy $E_{bh}$ have $f_{max}\approx \frac{E_{bh}}{3.266\varepsilon_{Pl}}+1$. Clearly, $f_{max}\geq 16$. Because $f(r)$ at the event horizon is 16. $f_{max}<16$ would mean the inner radius is greater than the event horizon. The minimum energy that a black hole can have is
\begin{equation}
	E_{bh, min}\approx 52.25\, \varepsilon_{Pl}
\end{equation}
This is the Planck energy Schwarzschild black hole (PESBH) where $f_{max}=f_0$.  
\section{Interaction}
The free field geometric quanta represent isolated geometries. I turn on the interaction to know what happens when fields interact with other fields.
\begin{equation}
	 \mathbf{H} = \mathbf{H}_0 + \mathbf{H}_{int}
\end{equation}
Here, the non-interacting Hamiltonian $\mathbf{H}_0 = \mathbf{H}_{\Phi}$ has explicit $\zeta$-dependence and $\mathbf{H}_{int}$ has interaction. 
The unitary operator is obtained using the following formula.
\begin{align*}
	&i\frac{d}{d\zeta} U\left(\zeta,\zeta^\prime\right) = \mathbf{H}_0 U\left(\zeta,\zeta^\prime\right);
	&U\left(\zeta,\zeta^\prime\right)= \left(\frac{\zeta}{\zeta^\prime}\right)^{-i\sqrt{\frac{32}{3}}\mathbf{n}}
\end{align*}
The evolution of the quantum state in interaction picture is given by, 
\begin{align*}
	&i\frac{d}{d\zeta} |\psi_\mathbf{n}(\zeta)\rangle_I = \mathbf{H}_I |\psi_\mathbf{n}(\zeta)\rangle_I \\
	& \mathbf{H}_I = \left(\frac{\zeta}{\zeta_0}\right)^{i\sqrt{\frac{32}{3}}\mathbf{n}} \mathbf{H}_{int} \left(\frac{\zeta}{\zeta_0}\right)^{-i\sqrt{\frac{32}{3}}\mathbf{n}}  
\end{align*}
I obtain the S-matrix from the unitary operator in the interaction picture as,
\begin{equation}
	S \coloneqq U_I \left(\zeta_0,-\infty\right) = T \exp{\left[-i \int_{\zeta_0}^\infty d\zeta\, \mathbf{H}_I\right]}
\end{equation}
$|\langle 0|\hat{S}|0\rangle |^2 \neq 1$ implies production of the geometric quantum from vacuum due to the presence of source. \\
\hspace*{3mm}Let us consider two free fields interacting in the following way.
\begin{equation}
	\mathbf{H}= \mathbf{H}_1 +\mathbf{H}_2 + \frac{\lambda}{2} \left(\Phi_1 - \Phi_2\right)^2
\end{equation}
In the large $\zeta$ limit, we can write the total Hamiltonian as a sum of two normal mode fields $\left(\Phi_+, \Phi_- \right)$. These normal mode fields are obtained by rotating $\left(\Phi_1, \Phi_2 \right)$ by an angle $\alpha =\pm\arctan \sqrt{\frac{2 \lambda +\omega _1^2-\omega _2^2}{\lambda}}$. Then normal mode frequencies are obtained as
\begin{align*}
	&\omega_{+} = \sqrt{\frac{\omega _1^2+\omega _2^2}{2}}
	&\omega_{-} = \sqrt{\frac{4 \lambda +\omega _1^2+\omega _2^2}{2}}
\end{align*}
The antisymmetric state has higher energy than the symmetric state. Without coupling, quanta are as good as individual Universes. If we take weak gravity limit of the Schwarzschild metric, i.e., $f\approx 1+\frac{\phi_{grav}}{r}$. The coupling effectively increases the value of $f(r)$. In other words, the radius is lesser than the sum of two non-interacting geometric quanta. The coupling gives the measure of distance between quanta. Stronger the coupling lesser will be the distance between them. 
\section{Results}
\hspace*{3mm}I have shown that the field satisfied by the Wheeler-DeWitt equation (\ref{FieldEqn}) cannot be a pure gravitational field. It can neither be a theory of multiverse nor a scalar theory of gravity. In addition to having geometric properties, fields satisfying the Wheeler-DeWitt equation have properties similar to the corresponding matter fields.\\
\hspace*{3mm}The field with positive intrinsic spatial curvature scalar $R^{(3)}$ is necessarily self-interactive. But the field with negative intrinsic spatial curvature scalar $R^{(3)}$ is not necessarily self-interactive. Fields with $R^{(3)}<0$ do not have interaction other than the geometric one as discussed in the \ref{features}. The self-interactiveness of fields depends upon the sign of $R^{(3)}$.\\
\hspace*{3mm}I observe that the gravitational field variables $( q_{ab}, P^{cd})$ and matter fields are not on equal footings. The gravitational field variables are first quantized. But matter fields are second quantized. Any theory that treats gravity and matter fields on equal footing concerning quantum level is dubious.\\
\hspace*{3mm}The re-interpretation geometrizes the quantum theory itself. The field $\Phi$ has the geometric quantum. There is no quantum corresponding to the pure gravitational field.\\
\hspace*{3mm}The creation and annihilation operators follow deformed algebra. $\beta$ is a function over superspace and not a constant. Since the role of creation and annihilation operators change depending upon the sign of $\beta$, the coherent state loses the coherence in the transition $\beta > 0 \leftrightarrow \beta < 0$.\\   
\hspace*{3mm} The theory dynamically resolves the big bang singularity and the black hole singularity. The Universe begins at finite minimum time $\zeta_{min}$. The ADM interpretation is that the Universe has non-zero initial volume. The big bang resolution is distinct from the loop quantum cosmology \citep{Kinjal} where there is the quantum big bounce. Initially, the 3-metric is exclusively time-dependent, and the non-trivial local geometries arise dynamically.\\
\hspace*{3mm}In the case of the Schwarzschild black hole, there exists an upper limit on the value of the 3-metric. The upper limit depends on the total energy of a given black hole. Since $\sqrt{f}$ is a dimensionless length of a black hole, the area of the black hole gets quantized.
\section{Conclusion}
\hspace*{3mm}The field $\Phi\left(\zeta, \zeta^A\right)$ describes the geometric matter. It satisfies the ADM constraints for \textit{pure gravity}. It also shows the properties that are close to the standard matter fields. Unlike pure gravity, the field has a non-trivial stress-energy tensor and, unlike matter fields that live in a spacetime, $\Phi$ itself is geometric. The field has a consistent asymptotically flat limit. Its single-geometric interpretation recovers the ADM theory.\\
\hspace*{3mm}The single geometric interpretation $\Phi\sim e^{\pm i P^{ab}q_{ab}}$ indicates that the higher-order couplings correspond to the matter fields. In the Higgs field as well, $\phi^4$-coupling gives mass. In the Klein-Gordon theory, the negative quadratic coupling constant is mathematical possibility only. The geometric field, on the contrary, has purely geometric intrinsic quadratic coupling. The $h$-coupling is not necessarily geometric. It is responsible for interpretation as geometric matter. This interpretation neither contradicts standard field theory nor gravity.\\
\hspace*{3mm}The quantum theory does not give the quantum of gravity. Instead, it gives the geometric quantum corresponding to a particular field. Non-linearity of the theory makes the creation and annihilation operators vector-valued in the spatial part of the supermetric. There exist a domain where the role of creation operator changes depending on the signature of (\ref{commutator}). In such case, coherent state loses it's coherence.\\
\hspace*{3mm}The theory resolves classical singularities dynamically by modifying the 3-metric near a singularity. In the case of the Schwarzschild geometry, an object falling inside cannot reach the center. It can approach only $f_{max}$. In the case of FLRW geometry, the Universe began at $\zeta_0\neq 0$ time. Even if we start with 3-metric $q_{ab}(t)$, the quantum dynamics inevitably introduces $q_{ab}(t,r)$.\\
\hspace*{3mm}The free field quanta are isolated geometries. There is no measure of the distance between non-interacting geometric quanta. In reality, fields interact with each other and give a sense of closeness. The geometric coupling $\lambda$ has dimensions of $\mu^2$-coupling. Closer geometric quanta have stronger geometric coupling between them. $h$-coupling has different units, and therefore, it has a different interpretation.\\
\hspace*{3mm}During the Planck epoch of the very early Universe, the energy of the geometric quanta is of the order of the Planck energy. Therefore, during this time, PESBHs get created. Beyond the Planck domain, PESBHs cannot form. These PESBHs undergo mergers and Hawking evaporation. The observation of primordial black holes would mean the existence of PESBH. This way, the theory expects the primordial black holes.\\
\hspace*{3mm}Resolution of classical singularities are quantum gravity effects, and situations such as the existence of gravitational decoherence show that the gravitational principles affect the quantum theory. In this sense, the quantum theory and gravity both receive modifications.
\def\bibsection{\section{\refname}} 
\bibliography{references}

\begin{thebibliography}{9}
\providecommand{\natexlab}[1]{#1}
\providecommand{\url}[1]{\texttt{#1}}
\expandafter\ifx\csname urlstyle\endcsname\relax
  \providecommand{\doi}[1]{doi: #1}\else
  \providecommand{\doi}{doi: \begingroup \urlstyle{rm}\Url}\fi

\bibitem[McGuigan(1988)]{ThirdQuantization}
Michael McGuigan.
\newblock Third quantization and wheeler-dewitt equation.
\newblock \emph{Physical Review D}, 38, 1988.

\bibitem[Kucher(2011)]{kucher}
Karel~V. Kucher.
\newblock Time and interpretations of quantum gravity.
\newblock \emph{International Journal of Modern Physics D}, 20\penalty0 (Suppl.
  1), 2011.
\newblock URL \url{https://doi.org/10.1142/S0218271811019347}.

\bibitem[Kiefer(2012)]{KieferBook}
Claus Kiefer.
\newblock \emph{Quantum Gravity 3rd edition}.
\newblock Oxford press, Oxford, UK, 2012.

\bibitem[Penrose(2014)]{penrose}
Roger Penrose.
\newblock On the gravitization of quantum mechanics 1: Quantum state reduction.
\newblock \emph{Foundations of Physics}, 44, 2014.
\newblock URL \url{https://doi.org/10.1007/s10701-013-9770-0}.

\bibitem[Cifford(1870)]{clifford}
W.~K. Cifford.
\newblock General theory of relativity - on the space theory of matter.
\newblock 1870.
\newblock URL \url{htttps://doi.org/10.1016/b978-0-08-017639-0.50009-1}.

\bibitem[Sten F.~Odenwald(1990)]{odenwald}
Zygon Sten F.~Odenwald.
\newblock A modern look at the origin of the universe.
\newblock \emph{Journal of Science and Religion}, 25, 1990.
\newblock URL \url{https://doi.org/10.1111/j.1467-9744.1990.tb00868.x}.

\bibitem[DeWitt(1967)]{DeWitt}
Bryce~S. DeWitt.
\newblock Quantum theory of gravity i: The canonical theory.
\newblock \emph{Physical Review (Series I)}, 160, 1967.
\newblock URL \url{https://doi.org/10.1103/physrev.160.1113}.

\bibitem[Giulini(1995)]{Giulini}
Domenico Giulini.
\newblock What is the geometry of superspace ?
\newblock \emph{General Relativity and Gravitation}, 51, 1995.
\newblock URL \url{https://doi.org/10.1103/PhysRevD.51.5630}.

\bibitem[Banerjee(2012)]{Kinjal}
Kinjal Banerjee.
\newblock Introduction to loop quantum cosmology.
\newblock \emph{Symmetry Integrability and Geometry Methods and Applications},
  2012.
\newblock URL \url{https://doi.org/10.3842/SIGMA.2012.016}.

\end{thebibliography}
\end{document}